\documentclass[prl,twocolumn,showpacs,preprintnumbers,amsmath,amssymb]{revtex4}
\usepackage{amsmath,graphicx}
\usepackage{amssymb}
\usepackage{pdfsync}
\def\bb{\textbf{b}}

\def\ff{\textbf{f}}

\def\pp{\textbf{p}}
\def\vv{\textbf{v}}

\def\rr{\textbf{r}}
\def\BB{\textbf{B}}

\def\EE{\textbf{E}}

\def\ss{\boldsymbol{\sigma}}

\def\f{\hat{f}}

\newcommand{\pder}[2]{{\frac{\partial #1}{\partial #2}}}

\makeatletter

\newcommand{\Rmnum}[1]{\expandafter\@slowromancap\romannumeral #1@}

\begin{document}
\title{Topological Magneto-Electric Effect Decay}
\author{D. A. Pesin}
\author{A. H. MacDonald}
\affiliation{Department of Physics, University of Texas at Austin, Austin TX 78712, USA}
\date{\today}
\begin{abstract}
We address the influence of realistic disorder on the effective magnetic monopole that is induced
near the surface of an ideal topological insulator (TI) by azimuthal currents which flow in response
to a suddenly introduced external electric charge.
We show that when the longitudinal conductivity $\sigma_{xx} = g (e^2/h)$ is accounted for,
the apparent position of a magnetic monopole initially retreats from the TI surface at speed
$v_{M} = \alpha c g$, where $\alpha$ is the fine structure constant and $c$ is the speed of light.
For the particular case of TI surface states described by a massive Dirac model, we further find that
the temperature $T=0$
Hall currents vanish once the surface charge has been redistributed to screen the external potential.
\end{abstract}
\maketitle
\noindent
\textit{Introduction--} When a time-reversal-symmetry breaking perturbation opens a
gap in the surface state spectrum of a three-dimensional topological insulator (TI)\cite{HasanKaneRMP,QiZhangRMP},
surface Hall currents and orbital magnetism are induced by
electrical perturbations.  This magneto-electric coupling effect can be
attractively described\cite{QiHughesZhang} by
adding a $ E \cdot B$ term to the electromagnetic Lagrangian.
The duality of the resulting {\em axion electrodynamics} model\cite{Wilczek}
leads to a curious {\em topological magneto-electric}
effect\cite{QiZhang,Vanderbilt,JamesII} in which an electric charge placed above the TI surface induces
Hall currents and associated orbital magnetization that appears to emanate from a magnetic monopole
below the surface.

In this paper we show that a non-zero TI surface state longitudinal conductivity
$\sigma_{xx} = g (e^2/h)$, an omnipresent experimental reality that
is not captured by the axion electrodynamics model, qualitatively alters the
topological magneto-electric effect.  We find that when the external charge is placed more
than a screening length $\lambda$ from the surface, the monopole moves away
with velocity $v_{M}=\alpha c g$.  In the long-time limit the screened external potential
becomes static.  In this case we find that the orbital magnetization response depends on details of the
surface state electronic structure, and that it vanishes in the particular case of a two-dimensional massive Dirac model
with temperature $T=0$ and a Fermi level position outside the gap.

\noindent
\textit{Macroscopic Theory--}
We assume here
that the TI surface has a well defined surface Hall conductivity and diffusion constant; this
assumption can fail for very well developed quantum Hall effects.
We first consider the limit in which the
separation $d$ between the external charge and the TI surface is larger than the screening length $\lambda$.
We introduce an external charge $Qe$ located a distance $d$ from the TI surface; since we wish to
treat this object as a source of macroscopic inhomogeneity rather than as a contribution to the
disorder potential we imagine that $Q \gg 1$ and that $d$ is longer than microscopic lengths.
Currents flow in the TI surface in response to the electric fields from the external charge and the
screening charges that accumulate in the TI surface layer. Working in two-dimensional momentum space and assuming that the total electric field changes sufficiently slowly with time, we use the continuity equation to conclude that
\begin{equation}
\label{eq:continuity}
 \pder{n^{2D}_q}{t}  = -
2 \pi \sigma_{xx} q \, (Q \exp(-qd) + n^{2D}_{q}).
\end{equation}
where $d$ is the distance from the surface to the external positive charge $Qe$ and
$n^{2D}_{q}$ is the Fourier transform of the induced surface state density.
In Eq.(\ref{eq:continuity}) we neglect the diffusion current, which is permissible at long distances
as we show below.   If we assume that the external charge is introduced
suddenly at time $t=0$ and that the two-dimensional (2D) density evolves in time in
accordance with Eq.~(\ref{eq:continuity}) we find that
\begin{equation}
n^{2D}_{q} = -Q \exp(-qd) (1 - \exp(-qv_M t))
\end{equation}
and that the total potential from external and screening charges is
\begin{equation}\label{eq:result for the potential}
 \phi_{tot}(q,t) = \frac{ 2 \pi eQ}{q} \; \exp(-q(d+v_Mt)).
\end{equation}
Here the monopole velocity $v_M = 2 \pi \sigma_{xx} = \alpha c g$ is large unless the
dissipative conductivity is much smaller than the quantum unit of conductance, {\it i.e.} unless
the quantum Hall effect on the TI surface is very well developed.
The potential at time $t$, which controls the instantaneous Hall currents and hence the
instantaneous magnetization is identical to that from a external charge that is located not
at vertical position $d$, but at vertical position $d+v_{M}t$.  As shown elsewhere\cite{QiZhang}, because of the
magneto-electric duality of axion electrodynamics, these Hall currents give rise to
a magnetization that is identical to that produced by a magnetic monopole located at a distance
$d+v_{M} t$ below the TI surface. Currents flow until macroscopic electric fields vanish.
The topological magneto-electric effect is therefore purely transient in the $d\gg \lambda$ limit.

\noindent
\textit{Screening in the Quantum Hall Regime--}
This result can be extended by including the diffusion contribution to the surface current:
\begin{equation}\label{eq:density_dynamics}
  \pder{n^{2D}_q}{t} = - 2\pi\sigma_{xx}q(Qe^{-qd}+n^{2D}_q) - D_Fq^2n^{2D}_q .
\end{equation}
The longitudinal conductivity, $\sigma_{xx}$ is related to the diffusion coefficient via the usual Einstein relation $\sigma_{xx}
= (\partial n/ \partial \mu) e^2D_F$.
Solving Eq.~(\ref{eq:density_dynamics}), we obtain the final expression for the total electric potential on the surface:
\begin{equation}\label{eq:final_potential}
 \phi_{tot}(q,t)= 2\pi eQ e^{-qd}\left(\frac{1+ (q\lambda_{TF})^{-1}
 e^{-(D_Fq^2+2\pi\sigma_{xx}q)t}}{q+ \lambda_{TF}^{-1}} \right),
\end{equation}
where $\lambda_{TF}^{-1} =2\pi \sigma_{xx}/D_F=2\pi \nu_F e^2$ is the screening
wavevector and $\lambda_{TF}$ the screening length.
The longitudinal currents vanish for $t \to \infty$ due to the Einstein-relation
cancellation between drift and diffusion contributions.
The total potential for $t \to \infty$ reduces to the standard result for
Thomas-Fermi screening in 2D.
Because $\partial n/ \partial \mu$ becomes extremely small when the
quantum Hall effect is well developed, $\lambda$ can be much
larger than typical microscopic length scales.

Since the external potential remains large for $t \to \infty$ at length scales smaller
than $\lambda$, there will be a
macroscopic orbital magnetic response to the screened potential
if the contributions to the Hall current from the screened electric field and
from the induced density inhomogeneities do not cancel.
Is there an Einstein relation for Hall currents?
Below we use a quantum kinetic theory to answer this question microscopically.
We conclude that the answer is no in general. Both drift and diffusion type terms do appear.
The contribution to the Hall current from density inhomogeneities can be
understood as being due to non-uniform internal magnetic moment~\cite{NiuRMP} densities.
Moreover, for the two-dimensional massive
Dirac equation that is normally used to model TI surface states, the drift and
diffusion Hall currents do cancel in the clean limit.

\noindent
\textit{Microscopic Theory--}
In the presence of an external potential the surface states of a 3D strong topological insulators can be
described\cite{HasanKaneRMP,QiZhangRMP} approximately by a 2D massive Dirac Hamiltonian:
\begin{eqnarray}\label{eq:s-hamiltonian}
  H_{{\rm{MD}}}=\int d^{2} \rr \, \Psi^\dagger\left(\BB_\pp\ss+e\phi_{ext}+U_{\rm{dis}}\right)\Psi.
\end{eqnarray}
Here $\BB_{\pp}=(vp_x,vp_y,\Delta)$ is a $\pp$-dependent
effective Zeeman field which acts on spinful surface
electrons.  With this choice for $\BB_\pp$, the Pauli matrices correspond to spins rotated by $\pi/2$ around the $\hat z$ axis, which we have taken to be normal to the surface. The mass term $\Delta$ breaks time-reversal-symmetry and  is normally thought of as arising from proximity exchange coupling to an insulating ferromagnet. For definiteness and without loss of generality, we take $\Delta>0$. $U_{\rm{dis}}$ describes an
atomic scale disorder potential which we take to be created by short-range impurities with concentration $n_{imp}$: $U_{\rm{dis}}=\sum_{i}u\delta(\rr-\rr_i)$. From now on we work in the system of units with $\hbar=1$.

In order to address the transport properties of this model, we use a
quasiclassical kinetic equation for the electron density matrix, $\hat f$, which takes the form
\begin{widetext}
\begin{equation}\label{eq:kineq}
  \partial_t \f_\pp+\frac{1}{2}\left\{\partial_\pp(\BB_\pp \cdot \ss),\partial_\rr\f_\pp\right\} +i[\BB_\pp \cdot \ss,\f_\pp] +e\EE_{\textrm{tot}} \cdot \partial_\pp\f^{\textrm{eq}}_\pp=\hat{I}_{st}.
 \end{equation}
\end{widetext}
In the above equation $\EE_{tot}$ is the total electric field including both external and induced potential
contributions, and $\hat{I}_{st}$ is the collision integral.\cite{collision}
We allow for an imperfect quantum Hall effect by considering the case in which
carriers are present in at least one of the bands due either to doping or to finite temperature.

The distribution function can be
decomposed into scalar and vector pieces, $ \hat f_\pp=n_\pp+\ss \cdot \ff_\pp$,
and the vector $\ff_\pp$ further separated into contributions parallel and
perpendicular to $\BB_\pp$, $f^{\parallel}_\pp \bb_\pp$ and $\ff^\perp_\pp$.
($\bb_\pp$ is a unit vector in the direction of $\BB_\pp$.)  In this parameterization of
the density matrix $n_\pp$ and $f^\parallel_\pp$ specify valence and conduction band occupation
numbers and $\ff^\perp_\pp$ interband coherence.
The kinetic equation for the full density matrix can be separated into a set of equations
for these components.

The model's intraband response is entirely standard~\cite{AshcroftMermin}, except
that scattering on the Fermi surface is influenced by the inner product of the momentum-dependent
conduction band states.  For the conduction band we find that
\begin{widetext}
\begin{equation}\label{eq:kinec_conduction}
  \pder{f^c}{t}+\vv_\pp\nabla f^c+e\EE_{tot}\vv_\pp\pder{n_F(B_\pp)}{B_\pp}=-\pi n_{imp}u^2\int\frac{d^2p'}{(2\pi)^2}\delta(B_\pp-B_{\pp'})(1+\bb_\pp\bb_{\pp'})(f^c_\pp-f^c_{\pp'}),
\end{equation}
\end{widetext}
where $\vv_\pp=v^2\pp/B_\pp$ is the band velocity appropriate for the conduction band of Hamiltonian~(\ref{eq:s-hamiltonian}).
It follows that the longitudinal conductivity, $\sigma_{xx}$ is related to the diffusion coefficient via the usual Einstein relation $\sigma_{xx}=\nu_Fe^2D_F$, where $\nu_F=B_{\pp_F}/2\pi v^2$ is the density of states at the Fermi level.
The absence of a longitudinal current in equilibrium, assumed in the macroscopic theory,
then follows from the cancelation between the second (diffusion) and third (drift)
terms of the left-hand-side of Eq.~(\ref{eq:kinec_conduction}),
when $f^c$ is replaced by its equilibrium Fermi function value.
The diffusion coefficient $D_F=\textrm{v}_\pp^2\tau_{tr}/2$ with
\begin{equation}
  \tau_{tr}^{-1}=\frac{n_{imp}u^2}{4v^2}\frac{v^2p_F^2+4\Delta^2}{\sqrt{v^2p_F^2+\Delta^2}}.
\end{equation}

\noindent
\textit{Hall response --} We have seen above that even in the presence of screening there is a residual
radially symmetric electric potential at the surface for $t\to\infty$.
The purpose of the following calculation is to determine whether or not that
potential can drive an azimuthal Hall current which contributes to the orbital magnetization.
The naive guess that one just has to multiply the screened electric field with the intrinsic Hall conductivity
to find the current fails because gradients in the density of carriers,
all of which generally carry intrinsic magnetic moments\cite{NiuRMP,orbitalmag}, also yield an azimuthal current.
The additional contribution can cancel the azimuthal
electric field response either completely, as it does in the longitudinal case, or partially.

Since the response we seek to evaluate includes the time-reveral-symmetry broken system's
anomalous Hall effect, we should include side-jump and skew scattering contributions\cite{Nagaosa_RMP} to
describe it fully in the presence of impurities. Since these are dependent on
impurity scattering at the Fermi surface, they can be obtained by considering
the leading quasiclassical corrections to Eq.~(\ref{eq:kinec_conduction}).
In the case of a uniform electric field, the quasiclassical kinetic equation for conduction band electrons
has been derived in Refs.~\cite{Luttinger,SinitsynDirac}.
This equation generalizes Eq.~(\ref{eq:kinec_conduction}) to include an
{\em anomalous distribution} generation term coming from the collision integral,
and beyond-Born-approximation skew scattering amplitudes.
Since we are interested here in response to a non-unform static electric field, we need to generalize the quasiclassical Boltzmann equation of Refs.~\cite{Luttinger,SinitsynDirac} to the non-uniform case by adding a drift term, $\tilde v_\pp\partial_\rr f^c$, just like the one in Eq.~(\ref{eq:kinec_conduction}), but with $\tilde v_\pp$ now including not only the band velocity, but also anomalous and side-jump corrections. It is then a simple matter to see that all electric-field drive terms vanish in that equation in local equilibrium. Therefore, side-jump and skew scattering contributions need not be considered and the entire Hall response comes from the intrinsic contribution.

The intrinsic contribution should be obtained from the equation for $\ff^\perp_\pp$.
Importantly, since we need not consider the side jump contribution, we can simply drop the contribution to the collision integral for $\ff^\perp_\pp$ coming from $f^\parallel_\pp$, since
 the latter gives a contribution to side-jump processes only.~\cite{Culcer}
Further, for a sufficiently clean surface, such that $B_\pp\tau_{tr}\gg 1$, we can also neglect the collisional relaxation of $\ff^\perp_\pp$ as compared to the precession term, coming from the commutator on the left hand side of
Eq.~(\ref{eq:kineq}).  The general expression for the static limit of $\ff^\perp_\pp$ is thus obtained
simply by isolating the inter-band terms on the left hand side  Eq.~(\ref{eq:kineq}).
We obtain
\begin{equation}\label{eq:kinec_perp}
  2B_\pp \ff^\perp_\pp=\left((\nabla n_\pp\partial_\pp)\BB_\pp\right)\times\bb_\pp+\left((e\EE_{tot}\partial_\pp)\ff_\pp\right)\times\bb_\pp.
\end{equation}
The second term on the right hand side of the above equation leads to the standard intrinsic contribution to the Hall conductivity due to the interband coherence created by the electric field.
The first term on the right hand side of Eq.~(\ref{eq:kinec_perp}) is the response to
the equilibrium density inhomogeneities.  Its contribution to the current can be seen to
equal the curl of the internal magnetic moment~\cite{NiuRMP} density of quasiparticles, which is nonuniform in space.
The right hand side of Eq.~(\ref{eq:kinec_perp}) does not necessarily vanish when
equilibrium values are used for $n_\pp = (f^c_{0\pp} + f^v_{0\pp})/2$ and
$\ff_\pp=(f^c_{0\pp} - f^v_{0\pp})/2$ ($f^{c,v}_0=n_F(\pm B_\pp-\mu+e\phi_{tot})$).
This property contrasts with Eq.~(\ref{eq:kinec_conduction}), in which both left and right hand sides vanish
under the local equilibrium ansatz.

Substituting the equilibrium values gives
$\nabla n_\pp=-e\EE_{tot}(\textrm{d} n_F(B_\pp-\mu)/\textrm{d}B_\pp-\textrm{d} n_F(-B_\pp-\mu)/\textrm{d}B_\pp)/2$ and
$\ff_\pp=\bb_\pp(n_F(B_\pp-\mu)-n_F(-B_\pp-\mu))/2$. Substituting these expressions in Eq.~(\ref{eq:kinec_perp}) and taking the local direction of the electric field to be along the $\hat x$ axis we obtain:
\begin{equation}
  \ff^\perp_\pp=-\frac{1}{4}eE_{tot}B_\pp(\partial_{p_x}\bb_\pp)\times\bb_\pp \sum_{\nu=\pm}\nu\pder{}{B_\pp}\left(\frac{n_F(\nu B_\pp-\mu)}{B_\pp}\right).
\end{equation}
This result will recover the usual intrinsic anomalous Hall conductivity when the derivative acts on the
$B_\pp^{-1}$ factor only.  When the derivative acts on both factors we obtain
\begin{widetext}
\begin{equation}\label{eq:jE}
\frac{j_y}{E_x} = \frac{e^2}{2}\int\frac{d^2p}{(2\pi)^2}\bb_\pp\cdot(\partial_{p_x}\bb_\pp)\times(\partial_{p_y}\bb_\pp) \sum_{\nu=\pm}\nu B_\pp^2\pder{}{B_\pp}\left(\frac{n_F(\nu B_\pp-\mu)}{B_\pp}\right).
\end{equation}
\end{widetext}
Note that the equilibrium value of $j_y/E_x$ is {\em not} the Hall conductivity.
The ratio instead describes equilibrium currents
that flow along equipotential lines of the screened external potential and
generate a contribution to the orbital magnetization.

\begin{figure}
\begin{center}
\includegraphics[width=3.0in]{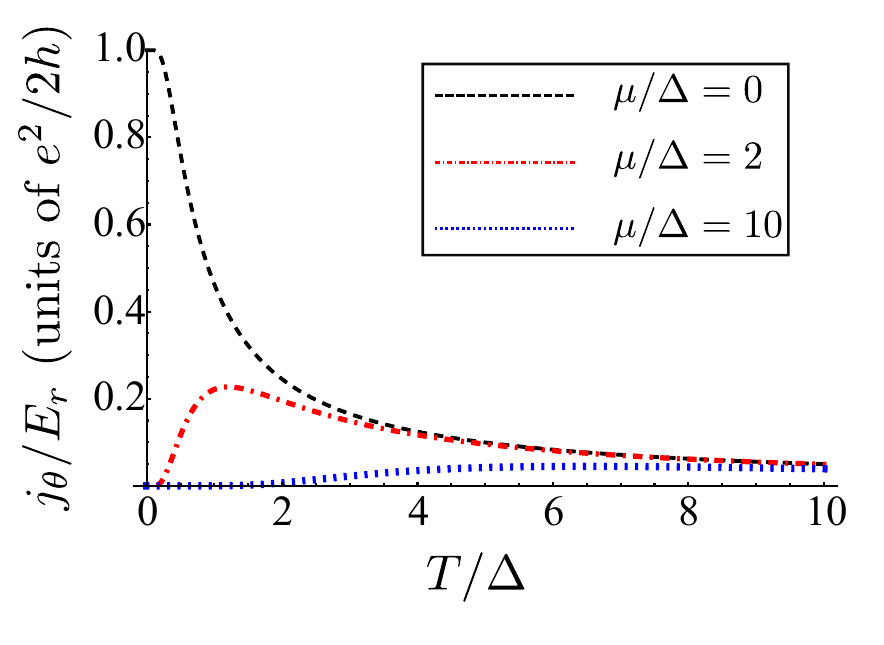}
\caption{(Color online) The dependence of the ratio of the azimuthal current, $j_\theta$, to the radial electric field, $E_r$, on temperature for different values of the chemical potential.}
\label{fig:finiteT}
\end{center}
\end{figure}

The right-hand-side of this expression vanishes for the 2D massive Dirac equation model for
temperature $T \to 0$.
In this special case Eq.~(\ref{eq:jE}) reduces to
\begin{equation}
  \frac{j_y}{E_x} = \frac{e^2}{4\pi}(n_F(-\Delta-\mu)-n_F(\Delta-\mu)),
\end{equation}
which vanishes for any $\mu>|\Delta|$ since the expression in brackets on the right hand side is the
Fermi factor difference between the top of the valence band and bottom of the conduction band.
Perfect cancelation occurs between the homogenous system anomalous Hall response and the current due to the curl of the internal quasiparticle magnetization density.
The same cancellation occurs for generalized Dirac models Eq.~(\ref{eq:s-hamiltonian}) with $|\pp|$-dependent velocities and constant $\Delta$, as long as the $\pp$-integrals are convergent.
This precise cancelation is however dependent on our
neglect of collisional relaxation in the equation for $\ff^\perp_\pp$, which would lead to $1/\Delta\tau_{tr}$ corrections.
The cancelation is also imperfect at finite temperature; substantial current signal can be recovered, as illustrated in Fig.~\ref{fig:finiteT}. The azimuthal current vanishes not only for
$T\to 0$ but also for $T\to \infty$ and is therefore a non-monotonic function of temperature.

\noindent
\textit{Discussion--}  When an external charge is placed near the surface of an ideal
TI with weak time-reversal symmetry breaking,
it induces an azimuthal current that produces\cite{QiZhang} the same magnetic field as would be
produced by a magnetic monopole located below the TI surface.
The axion electrodynamics model\cite{Wilczek,QiHughesZhang} of TI
magneto-electric and magneto-optical properties\cite{Karch,JamesII,Maciejko2010} elegantly captures
this intriguing property.  In this paper we have examined how the Hall current response
is altered by the samples imperfections which always results in a finite longitudinal conductivity $\sigma_{xx} = g (e^2/h)$.
Systems with a finite $\sigma_{xx}$ are not fully described\cite{JamesII,Nomura} by the axion electrodynamics model
so we develop our theory directly in terms of surface state electronic properties.
We find that the apparent monopole position moves away from the TI surface with a
velocity $v_{M} =\alpha c g$.  Since graphene based two-dimensional electron
systems, which are similar to TI surface states, can\cite{Tzalenchuk} have $g$ values $\sim 10^{-7}$ or smaller
when time-reversal symmetry is broken by
an external magnetic field, there is a reasonable hope  that it will
be possible to obtain TI samples in which $v_{M}$ is small enough to enable observations
in which $\sigma_{xx}$ plays no role and the axion electrodynamics model
is applicable. There is a considerable recent experimental effort in this direction.~\cite{Expt}

In the long-time limit after the external charge
screening process has been completed, we find that the azimuthal current response
has two contributions, one proportional to the Hall conductivity and
treated previously by Zang and Nagaosa,\cite{Zang} and one proportional to an
external potential induced change in the internal magnetization\cite{orbitalmag} of the surface states.
For the particular case of a massive Dirac model the
two contributions cancel exactly in the clean $T=0$ limit in the presence of a Fermi surface.
We obtain this result using a quasiclassical kinetic equation approach,
which may not be reliable near band edges due to both quantum and
non-linear screening effects, but nevertheless starkly demonstrates the
distinction between azimuthal current and Hall conductivity responses.
The special properties of the massive Dirac model are related to its well known~\cite{2DDiracOrbital}
unusual orbital magnetization properties in the uniform system limit.
In general the magnetic flux induced by an
electron charge near a time-reversal symmetry broken TI surface
is dependent on the $|\pp|$-dependence of the
exchange potential $\Delta_\pp$ and disorder effects,
and not simply on surface's  Hall conductivity.

\acknowledgements
The authors are grateful to Dimitrie Culcer, Alexey Kovalev, Qian Niu,
Nikolai Sinitsyn, and Boris Spivak for useful discussions.
This work has been supported by Welch Foundation grant TBF1473, NRI-SWAN, and DOE Division of
Materials Sciences and Engineering grant DE-FG03-02ER45958.

\end{document}